\documentclass[runningheads]{llncs}

\usepackage[T1]{fontenc}
\usepackage{graphicx}
\usepackage{amsmath}
\usepackage{hyperref}
\usepackage{here}

\usepackage[american]{babel}
\usepackage{microtype}

\usepackage{subcaption} 
\usepackage{float}
\usepackage{paralist}
\usepackage{wrapfig}
\usepackage{cleveref}
\usepackage{xspace}
\usepackage{mathtools}
\usepackage{hyperref}
\usepackage{array}
\usepackage{longtable}
\usepackage{todonotes}
\usepackage{geometry}

\definecolor{bleu}     {RGB}{ 49,140,231}

\hypersetup{
  colorlinks,
  urlcolor  = bleu,
  citecolor = bleu,
  linkcolor = bleu,
}

\newcommand{\FRC}  {\ensuremath{\kappa_{\text{FR}}\xspace}}
\newcommand{\ORC}  {\ensuremath{\kappa_{\text{OR}}\xspace}}
\newcommand{\PRN}{Physician Referral Networks\xspace}
\newcommand{\apparent}{\href{https://apparent.topology.rocks/}{\texttt{apparent}}}
\begin{document}

\title{Characterizing \PRN with Ricci Curvature}
\titlerunning{Referral Network Curvature}
%
\author{Jeremy Wayland \inst{1,2}\orcidID{0000-0002-8766-8737} \and
Russell J. Funk\inst{4}\orcidID{0000-0001-6670-4981} \and
Bastian Rieck\inst{1,2,3}\orcidID{0000-0003-4335-0302}}
\authorrunning{Wayland et al.}
%
\institute{Technical University of Munich
School of Computation, Information and Technology
Arcisstraße 21
80333 Munich
Germany \and Institute of AI for Health, Helmholtz Munich, Ingolstädter Landstraße 1, 85764 Oberschleißheim Germany \\
\email{jeremydon.wayland@helmholtz-munich.de}\\
\and
Department of Informatics, University of Fribourg, Boulevard de Pérolles
90, Fribourg, 1700, Fribourg, Switzerland \\
\email{bastian.grossenbacher@unifr.ch} \\
 \and
 Carlson School of Management
 University of Minnesota
 321 19th Avenue South
 Minneapolis, MN 55455
 USA\\
\email{rfunk@umn.edu}}
\maketitle              

\newcommand{\BR}[1]{\todo[%
  author    = BR,
  color     = lightgray,
  inline,
]{#1}}

\newcommand{\RF}[1]{\todo[%
  author    = RF,
  color     = yellow,
  inline,
]{#1}}

\begin{abstract}
    Identifying~(a) systemic barriers to quality healthcare access and~(b) key indicators of care efficacy in the United States remains a significant challenge. To improve our understanding of regional disparities in care delivery, we introduce a novel application of curvature, a geometrical-topological property of networks, to \PRN. Our initial findings reveal that \emph{Forman-Ricci} and \emph{Ollivier-Ricci} curvature measures, which are known for their expressive power in characterizing network structure, offer promising indicators for detecting variations in healthcare efficacy while capturing a range of significant regional demographic features. We also present \apparent, an open-source tool that leverages Ricci curvature and other network features to examine correlations between regional \PRN structure, local census data, healthcare effectiveness, and patient outcomes.
    \end{abstract}

\section{Introduction}

In the rapidly evolving field of healthcare management, the analysis of medical claims data has become an essential component for improving the quality and equity of healthcare services. The nature of care delivery in the United States is heavily influenced by its fragmentation---care is often spread across multiple disconnected providers (e.g., primary-care physicians, specialists). Settings with greater care fragmentation have been shown to inhibit effective communication and coordination between care team members, thus contributing to higher costs and lower quality of care~\cite{frandsen2015care,snow2020patient,juo2019care,agha2019fragmented,cebul2008organizational}.

Despite the well-understood impacts of fragmentation, there are still few quantitative tools that can capture the mechanisms of care delivery networks at scale \cite{funk2018association}. Standard analyses of local infrastructure features, often executed using tabular data, are limited in their ability to distill complex dynamics among physicians. Graph representations and tools from network analysis have thus been adopted in the literature to provide more sophisticated models for care delivery~\cite{hollingsworth2015differences,popescu2024segregation,matthews2023within,an2018analysis,an2018referral,barnett2012physician,casalino2015physician,everson2018repeated,funk2018association,kim2019informal,landon2018patient,pollack2013patient,casalino2015physician}, permitting an understanding of the data both on the local and the global level. 
Increased access to medical claims data has placed an emphasis on analyzing \emph{\PRN}, which encode patient sharing between physicians across different regions in the United States. \PRN are typically modeled using administrative claims data (often from Medicare) to map patient-flow patterns, elucidating more complex dynamics that underpin local care delivery. Their study has proved fruitful for enhancing care coordination, reducing costs, improving patient outcomes, and understanding social determinants of health~\cite{dugoff_scoping_2018,gandre_care_2020,ghomrawi_physician_2018,landon_variation_2012}.

Our work aims to introduce novel foundational tools to the study of \PRN. In particular, we focus on curvature-based measures for capturing structural properties of a graph. Ricci curvature, a well-known concept in differential geometry, has recently gained traction for various applications in the field of machine learning~\cite{southern_curvature_2023,coupette2023ollivierricci,Devriendt_2022}.
In fact, \emph{positive} Ricci curvature in a network was shown to correspond to regions of high connectivity–-–which, in the context of care delivery, could represent tighter-knit groups with the capacity for better coordination and communication. \emph{Negative} curvature, by contrast, is known to capture information bottlenecks in message-passing neural networks \cite{topping2022understanding}, and has the potential to indicate less robust regions in a referral network that are prone to information loss, miscommunication, and potentially lower standards of care. 

In addition to strong theoretical foundations and promising interpretations in the context of care delivery, Ricci curvature measures also stand out against standard network descriptors (e.g., assortativity, centrality, clustering coefficients) due to their ability to precisely describe the structure of a network and their flexibility in easily switching between edge-level, node-level, or network-level aggregations to suit the analysis at hand \cite{southern_expressive_2023}. Thus, we aim to bolster foundational analysis of \PRN with the following \textbf{key contributions}:

\begin{compactenum}
    \item We introduce Ricci curvature-based measures as novel structural features for characterizing \PRN.
    \item We perform initial analyses using these measures to study healthcare delivery across diverse U.S. regions. 
    \item We develop \apparent, an open-access tool that empowers researchers studying U.S. healthcare delivery by offering an interactive platform to explore referral network features and their correlations with local census data, healthcare effectiveness, and patient outcomes. 
\end{compactenum}

\section{Background}

Research on \PRN has evolved significantly, transitioning from small-scale empirical studies to large-scale analyses enabled by the advent of large-scale data access \cite{shortell1974determinants,shortell1971physician,javalgi1993physicians,landon_variation_2012,barnett2011mapping,funk2018association}. This progression has expanded the scope and depth of understanding regarding the structure and dynamics of care delivery networks.
Early empirical studies, despite being constrained to individual care delivery networks and often involving only several thousand physicians, provided essential insights into the conceptualization of \PRN and their ability to characterize healthcare delivery \cite{dugoff_scoping_2018,landon_variation_2012}. Recently, the growing availability of large-scale administrative data, particularly from electronic medical records and insurance claims, has allowed for more robust studies on \PRN across the country. This work examines quantitative network properties to glean insights on care coordination, utilization, and cost, as well as relationships between network structure and economic/clinical outcomes \cite{gandre_care_2020,hollingsworth2015differences,popescu2024segregation,everson2018repeated}.

However, significant challenges remain in effectively and efficiently extracting relevant features from these complex data structures. Key obstacles include the high dimensionality of network data, the computational complexity of analyzing large-scale networks, and the difficulty in interpreting network measures in clinical contexts.
This creates a crucial need for new analytical approaches.
Bridging the gap between foundational theory and contemporary measurement has already shown promising results for care delivery analysis \cite{funk2018association,kim2019informal,pollack2013patient}. By introducing Ricci curvature, we aim to further expand this frontier, leveraging a rich mathematical tradition in differential geometry to enhance our collective interpretations when evaluating these intricate healthcare systems.

\begin{table}[ht!]
    \centering
    \resizebox{\textwidth}{!}{
    \begin{tabular}{| m{5cm} | m{11cm} |}
    \hline
    \centering \textbf{Table Name} & \textbf{Description} \\ \hline
    \centering \textbf{referral\_network\_features} & Features of \PRN by HSA and year, including network metrics such as assortativity, clustering, density, Forman-Ricci, and Ollivier-Ricci Curvatures. These metrics provide insights into the structure and connectivity of local physician networks. \\ \hline
    \centering \textbf{hedis\_measures} & Contains Healthcare Effectiveness Data and Information Set (HEDIS) measures by Health Service Area (HSA) and year, including various diabetes and mammography metrics segmented by race. These measures are crucial for evaluating healthcare quality and outcomes across different regions and demographics. \\ \hline
    \centering \textbf{hospital\_atlas\_data} & Includes information about hospitals such as provider details, location, teaching status, and bed counts. This data is essential for understanding the distribution and capacity of healthcare facilities within local networks. \\ \hline
    \centering \textbf{population\_census} & Provides demographic data by HSA and year, including population counts by race, median household income, employment, and education levels. This demographic information is important for analyzing the socioeconomic context of healthcare networks. \\ \hline
    \centering \textbf{post\_discharge\_records} & Contains post-discharge outcomes for various conditions by HSA and year, such as readmission rates and follow-up care. This data helps assess the effectiveness of patient transitions from hospital to home care. \\ \hline
    \centering \textbf{standard\_pricing} & Includes standard pricing and payments for various medical services by HSA and year. Understanding pricing variations is important for economic analyses of healthcare delivery and cost efficiency. \\ \hline
    \centering \textbf{local\_physician\_interactions} & Contains Medicare claims data by year and HSA, detailing interactions between physicians (identified by NPI numbers). This data is the sole generator for the \PRN. \\ \hline
    \end{tabular}}
    \vspace*{0.2cm}
    \caption{\textbf{Data Release Summary}. We provide an overview of
    the tables that comprise our original database, along with high
  level descriptions of their features and relevance for characterizing
local \PRN. Up-to-date table structure, data summaries, and feature descriptions will be maintained within our prototype, \apparent.}
    \label{tab:database-summary}
\end{table}

\section{Data Description}

We analyze Medicare claims data in the United States from 2014--2017, consisting of 663,541 total physicians, with approximately 10,000,000 patient-sharing records per year across 3404 Hospital Service Areas (HSAs). These data were made publicly available by \href{https://www.careset.com/}{CareSet} a healthcare data consultancy, in collaboration with the Center for Medicare and Medicaid Services (CMS), the U.S. government agency that administers Medicare, and have been widely used in prior research on physician networks~\cite{gebhart2021go,everson2022electronic,graves2023physician,kim2023structure}. As outlined in \Cref{tab:database-summary},  we supplement physician interactions with data from several sources including basic data on the physicians included in our networks~(e.g., practice locations, specialty types) from the National Plan and Provider Enumeration System~(NPPES) as well as metrics on local healthcare quality and spending from the Dartmouth Institute for Health Policy and Clinical Practice. 

\paragraph{Data format.}
The data are provided in the form of edge lists (i.e., pairs of physicians). The nodes correspond to physicians, who are assigned unique, numerical codes corresponding to their National Provider Identifiers~(NPIs). Physicians are linked by an edge when they treat the same patients within a specified timeframe. The strength of this connection is measured by the number of patients they have in common. For instance, if Physician A treated 30 patients in one week, and 12 of those patients were seen by Physician B the following week, we would establish a connection between A and B with a strength value of 12. We exclude from our analysis organizational providers (e.g., hospitals, clinics) such that all nodes in our networks correspond to individuals. We also limit the networks to include only physicians who are likely to be directly involved with patient care~(e.g., we remove radiologists). The list of included physicians was developed in consultation with a team of clinicians. 

\paragraph{Location information.}
Healthcare delivery in the United States tends to be highly localized, with substantial variation in both healthcare practices and outcomes across regions. To capture this variation, we map physician networks separately by region, focusing on Hospital Service Areas (HSAs; as defined by the Dartmouth Atlas). HSAs have been widely used in prior research on healthcare delivery and are designed to correspond roughly to local markets for hospital services. We position physicians within HSAs by geocoding their practice locations; as such, our network maps consist of shared patients among the physicians practicing within a particular HSA. Finally, we study networks separately by year, for each of the four years in our study window, recognizing that the structure of \PRN may change over time. 

\paragraph{Selection criteria.}
To protect patient privacy and comply with CMS regulations, we only include physician pairs in our data if they shared at least 11 patients within a specific time period \cite{CMS2439F}. While this thresholding approach necessarily results in some data loss, previous research suggests that connections between physicians based on few shared patients are often not considered significant by the physicians themselves~\cite{barnett2011mapping}. The most meaningful professional relationships typically involve a higher number of shared patients and align more closely with survey data. Therefore, this required minimum threshold is actually beneficial, as it likely produces networks that more accurately reflect real-world professional relationships.

\section{Methods}

\subsection{Building \PRN}
Similar to social or biological networks, we can model \PRN as graphs. A graph $G \coloneqq (V,E)$ consists of a set of nodes $V$ that model the objects of interest in a system and a set of edges $E$ that encode relationships between the nodes. In the context of care delivery, nodes often represent healthcare providers, such as physicians, and edges can represent patient-sharing relationships or referrals between these providers.

\begin{figure}[ht]
    \centering
    \begin{subfigure}[b]{0.48\textwidth}
        \centering
        \includegraphics[width=0.8\textwidth]{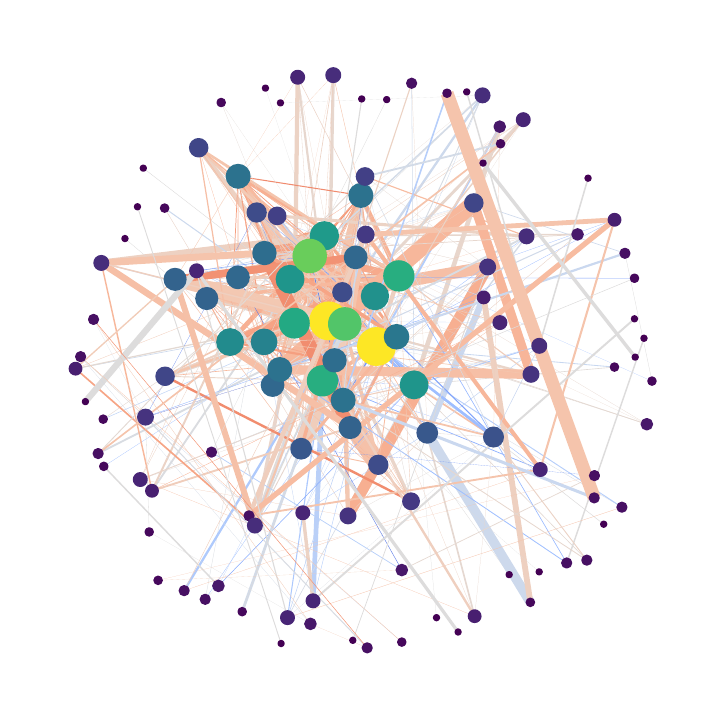}
        \caption{Santa Ana, CA}
        \label{fig:socal-network}
    \end{subfigure}
    \begin{subfigure}[b]{0.48\textwidth}
        \centering
        \includegraphics[width=0.8\textwidth]{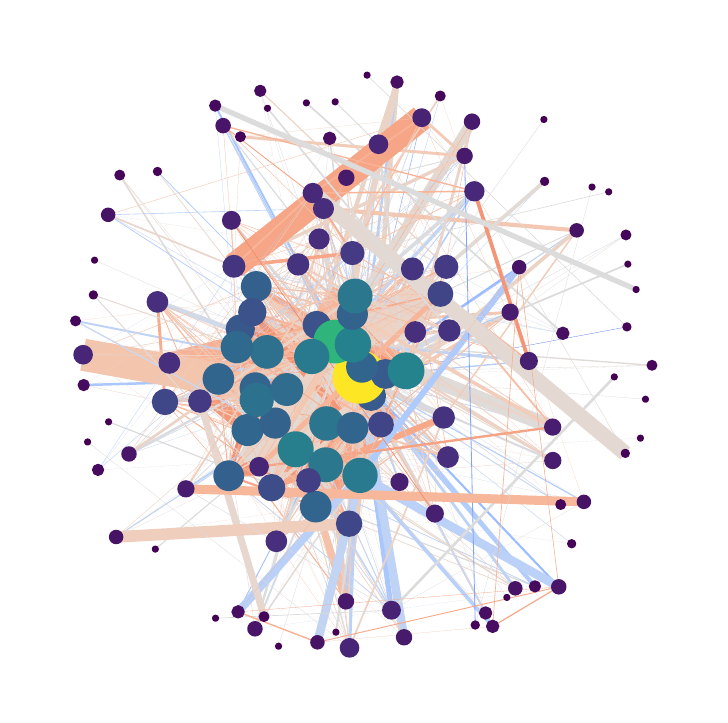}
        \caption{Snellville, GA}
        \label{fig:atlanta-network}
    \end{subfigure}
    \caption{ \textbf{Visualizing \PRN}. Above, we visualize networks from regions with known differences in care delivery. \Cref{fig:socal-network} shows a referral network from Santa Ana, California, exhibiting a higher quality of care, while \Cref{fig:atlanta-network} comes from Snellville, Georgia. As made available in \apparent, we combine traditional network features with Ricci Curvature to elucidate the structural properties of care delivery in each region: Nodes are sized and colored by centrality and degree while edges are sized and colored by number of referrals and \ORC, respectively. 
    \textit{Interpretation:} Larger yellow nodes are physicians that play a prominent role in regional care delivery by dictating patient flow. Blue-gray edges (negative curvature) are potential bottlenecks for patients finding the best provider, while orange edges (positive curvature) indicate strong communication between physicians and well-established patient flow.}
    \label{fig:real-networks}
\end{figure}

Our goal is to capture patient flow within an HSA during the course of a year. Each provider in the HSA becomes a node in the graph, and edges are assigned between physicians when a sufficient number of patients are shared via referrals. We provide some real examples in \Cref{fig:real-networks} that arise from HSAs in different regions of the country (Southern California vs. Atlanta). Our graph representations are built directly from the \texttt{local\_physician\_interactions} table (see \Cref{tab:database-summary}).

\subsection{Ricci Curvature}

\begin{figure}[ht!]
    \centering
    \begin{subfigure}{0.4\textwidth}
        \centering
        \includegraphics[width=1.2\textwidth]{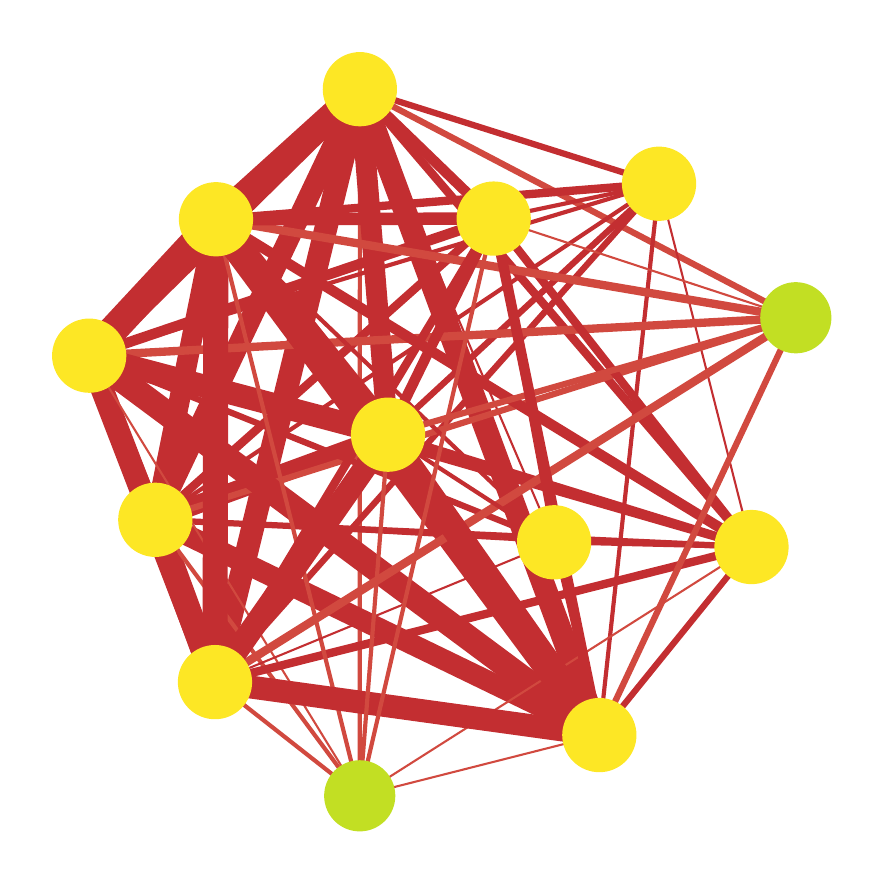}
        \caption{High $\ORC$}
        \label{fig:max_curvature}
    \end{subfigure}
    \hspace{0.1\textwidth}
    \begin{subfigure}{0.4\textwidth}
        \centering
        \includegraphics[width=1.2\textwidth]{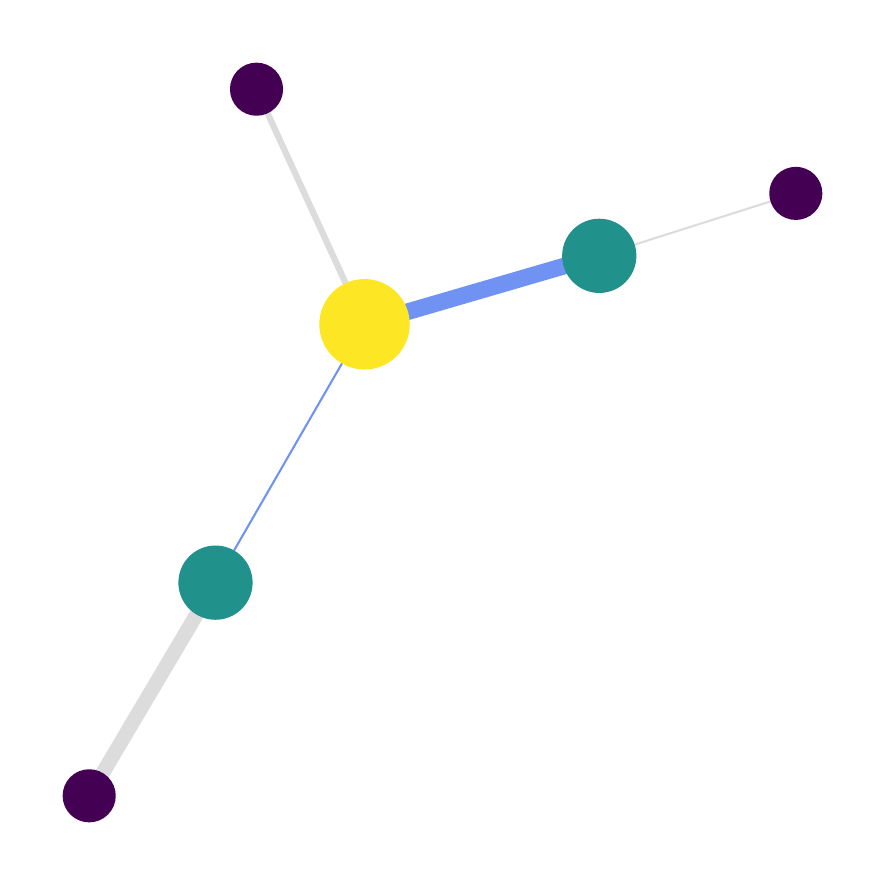}
        \caption{Low $\ORC$}
        \label{fig:min_curvature}
    \end{subfigure}
    \caption{\textbf{Ollivier-Ricci Curvature for \PRN}. In care delivery networks, \emph{positive} Ricci curvature, as seen in \Cref{fig:max_curvature}, can highlight regions with high connectivity and well-coordinated groups, potentially indicating better communication and care coordination. Conversely, \emph{negative} curvature, seen in \Cref{fig:min_curvature}, can signal areas prone to information bottlenecks, which might lead to communication issues and lower care quality.}
    \label{fig:curvature-examples}
\end{figure}

Ricci curvature is a fundamental concept situated at the intersection of differential geometry and topology that has recently emerged as a powerful tool in graph machine learning~\cite{topping2022understanding,coupette2023ollivierricci,Devriendt_2022,ni_ricci_2015,sia_ollivier-ricci_2019}. Curvature measures for graphs offer a concise yet comprehensive way to represent and analyze structural features of networks, serving as a lens for measuring local ``cohesiveness.'' Among the various curvature constructions, our work focuses on two specific types of Ricci curvature: \emph{Forman-Ricci} and \emph{Ollivier-Ricci}. Both of these measures evaluate how the volume growth of a network deviates from that of a ``model'' Euclidean space, providing nuanced insights into network structure. This is achieved by capturing the similarity between neighborhoods of nodes. For some token examples of positively- and negatively-curved \PRN, see \Cref{fig:curvature-examples} 

Similar to their continuous counterparts, discrete network curvature measures permit the following interpretations:
\begin{compactitem}
\item \textbf{Positive curvature} indicates configurations with overlapping neighborhoods, such as cliques, where nodes are well-connected.
\item \textbf{Zero or low curvature} is associated with grid-like structures, where connectivity is more uniform but less dense. 
\item \textbf{Negative curvature} often appears in tree-like formations, where nodes have fewer overlapping neighbors and connectivity is more sparse.
\end{compactitem}
In comparison with standard network descriptors such as assortativity, centrality, and clustering coefficients, Ricci curvature offers greater \emph{expressiveness}---meaning these measures are more effective at distinguishing between non-isomorphic graphs~\cite{southern_expressive_2023}. Additionally, curvature measures can be aggregated to provide insights at different levels (edge, node, or network) allowing for a detailed analysis of specific physicians, their relationships, or a summary comparison between different networks. We hope to inspire practitioners and researchers interested in healthcare management and delivery systems to incorporate curvature-based measures into their network analysis pipelines.

\subsubsection{Forman-Ricci Curvature}
Forman-Ricci Curvature (FRC) is a discrete curvature measure for edges in a network based on local connectivity. In the context of physician networks, FRC measure provides an efficient, albeit less expressive, measure for evaluating the collaboration patterns among physicians. The formula for Forman-Ricci Curvature between two connected nodes $i$ and $j$ is given by:
\begin{equation} \label{eq:forman}
\FRC(i,j) \coloneqq 4 - d_i - d_j + 3|\#_{\Delta}|
\end{equation}
Here, $d_i$ and $d_j$ denote the degrees of nodes $i$ and $j$, respectively, and $|\#_{\Delta}|$ represents the number of triangles (i.e., 3-cliques) that include the edge $(i,j)$. In \PRN, $d_i$ and $d_j$ correspond to the number of direct connections (referrals) each physician has, while $|\#_{\Delta}|$ indicates the number of mutual connections between the two physicians. A higher FRC value suggests a stronger collaborative relationship, characterized by more mutual connections and shared professional interactions. In general FRC values are unbounded---in large networks edges can take large positive or large negative values (see \Cref{fig:forman-distribution}). As an intuitive example for an edge with very \textit{negative} $\FRC$, imagine two communities modeled as fully connected graphs that are connected by a single edge. Each community can be made to have arbitrarily large degrees ($d_i,d_j$ in \Cref{eq:forman}), but the edge between them cannot have mutual connections (i.e., $|\#_{\Delta}|=0$).

\begin{figure}[ht!]
    \centering
    \includegraphics[width=\textwidth]{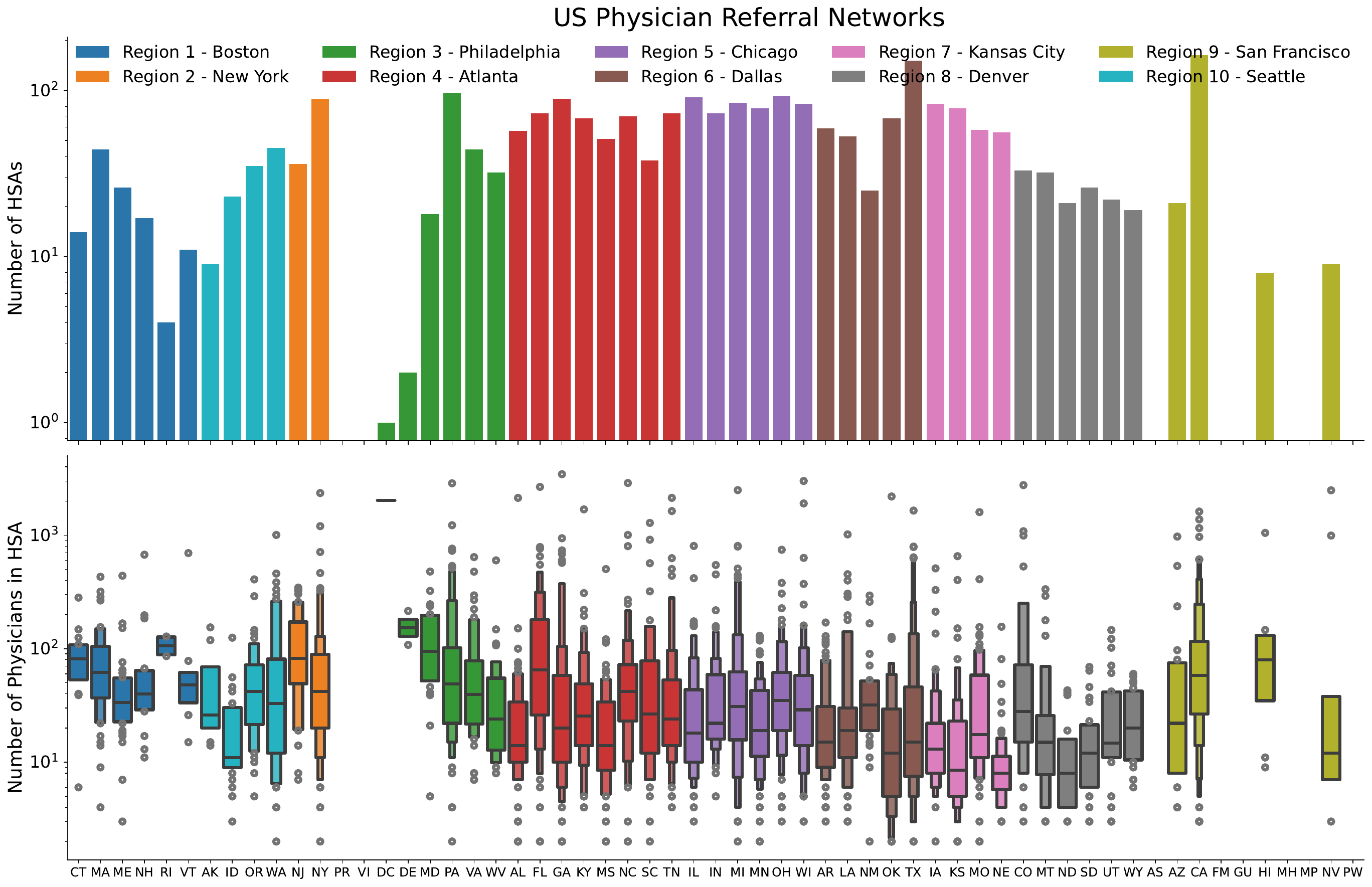}
    \caption{\textbf{Regional Distribution of \PRN}. Here we visualize the distribution of \PRN in 2017 (one per Health Service Area, HSA) alongside physician counts (nodes) by state, color-coded by region. This figure highlights the relative \emph{size} and \emph{density} of \PRN and establishes a baseline for structural differences in care coordination across the United States.}
    \label{fig:network-distribution}
\end{figure}

\subsubsection{Ollivier-Ricci Curvature}
Ollivier-Ricci Curvature (ORC) is a more expressive measure than FRC that quantifies the curvature of edges in a graph by comparing the neighborhoods of the nodes connected by the edge. Rather than counting triangles, a more sophisticated comparison is achieved using methods from optimal-transport theory; specifically, the Wasserstein distance~(also known as Earth Mover's distance) between the probability distributions associated with node neighborhoods is evaluated.
The formula for Ollivier-Ricci Curvature between two connected nodes $i$ and $j$ is given by:
\begin{equation}  \label{eq:orc}
\ORC(i,j) \coloneqq 1 - \frac{1}{d_G(i, j)} W_{1}(\mu_{i}, \mu_{j})
\end{equation}
Here, $d_G(i, j)$ represents the shortest path distance between nodes $i$ and $j$ in the graph, and $W_{1}(\mu_{i}, \mu_{j})$ denotes the Wasserstein distance between the probability measures $\mu_{i}$ and $\mu_{j}$, which are defined over the neighborhoods of nodes $i$ and $j$, respectively.\footnote{Our computations use uniform probability distributions.} 

In the context of physician networks, $\mu_{i}$ and $\mu_{j}$ can be understood as distributions of patient referrals or collaborations within the neighborhoods of physicians $i$ and $j$. A higher ORC value indicates that the neighborhoods of the two nodes are more similar, suggesting more cohesive collaboration patterns. For extreme examples or positive and negative ORC, see \Cref{fig:curvature-examples}. 
Despite its enhanced expressiveness in capturing distinct structural properties of networks, even for challenging tasks such as distinguishing Strongly-Regular and Rook-Shrikhande graphs \cite{southern_expressive_2023}, ORC is computationally expensive for very large graphs. In contrast, FRC offers a more scalable alternative. If FRC can capture meaningful signals, it provides a practical advantage as it can be computed efficiently, even for the largest \PRN in the United States.

\section{Results}

Our interactive database contains information on approximately 13,500 physician networks (bounded by HSAs), sourced over the years 2014--2017. Along with approximately 20 network descriptors, including statistics for network assortativity, clustering coefficients, density, centrality, Forman-Ricci Curvature, and Ollivier-Ricci Curvature, we also provide infrastructure for querying rich metadata for these networks. \Cref{tab:database-summary} provides a summary of the available tables and their basic descriptions.

As a preliminary case study that highlights the utility of our proposed techniques, we perform an analysis of the regional referral networks.
More specifically, we explore approximately 4000 HSAs with available referral data in 2017. \Cref{fig:network-distribution} describes the distribution of \PRN by state, split into 10 larger regions often used for healthcare analysis, while also depicting the distribution of network sizes.

\begin{figure}[ht!]
    \centering
        \includegraphics[width=\textwidth]{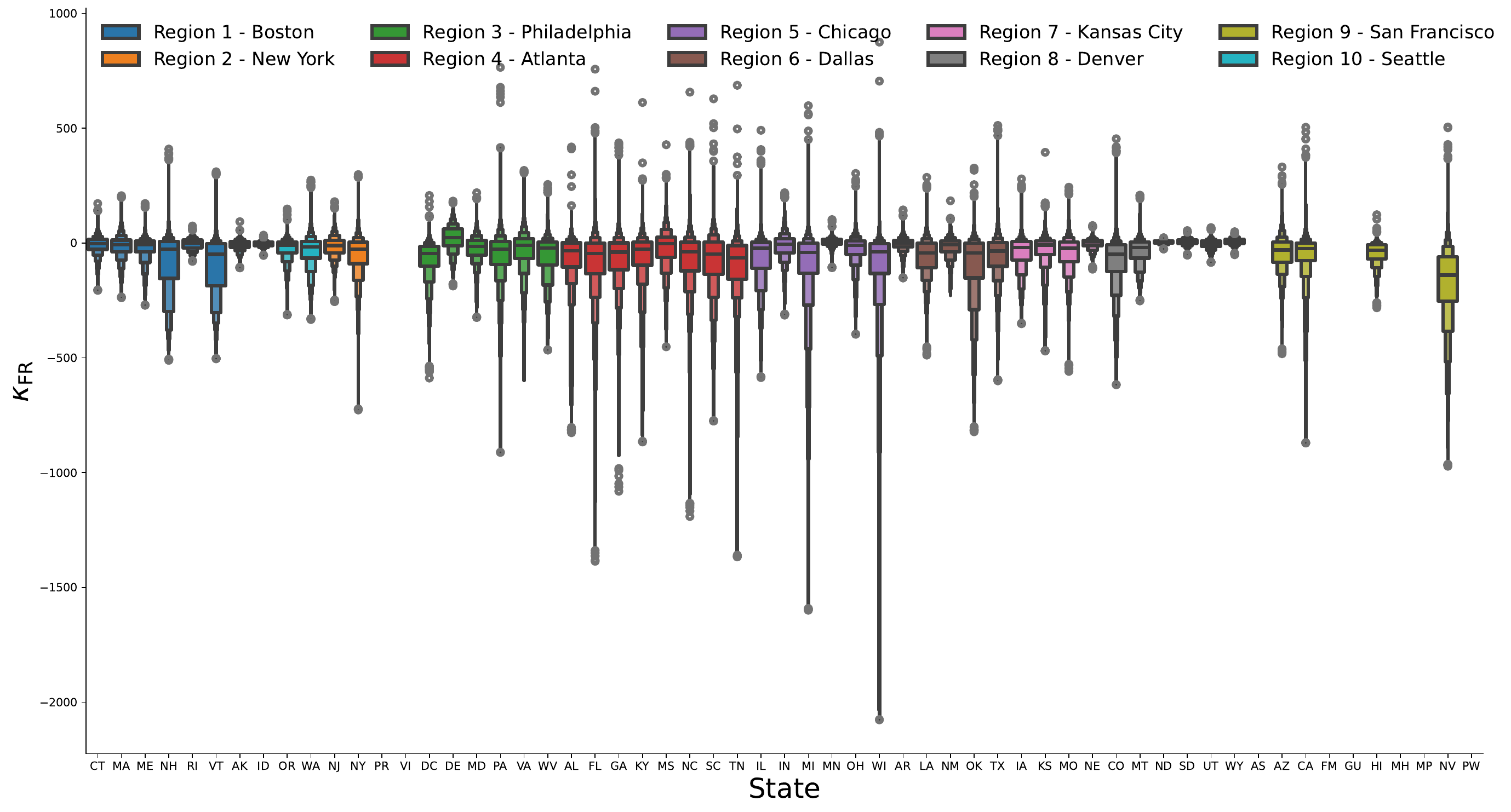}
        \caption{\textbf{Regional Distribution of Forman-Ricci Curvature.} This figure displays a wide range of inter- and intra-regional differences in care delivery structure as measured by $\FRC$. We plot curvature values for each edge (provider connection) for all 2017 \PRN. }
        \label{fig:forman-distribution}
\end{figure}

\begin{figure}[hb!]
    \centering
    \includegraphics[width=\textwidth]{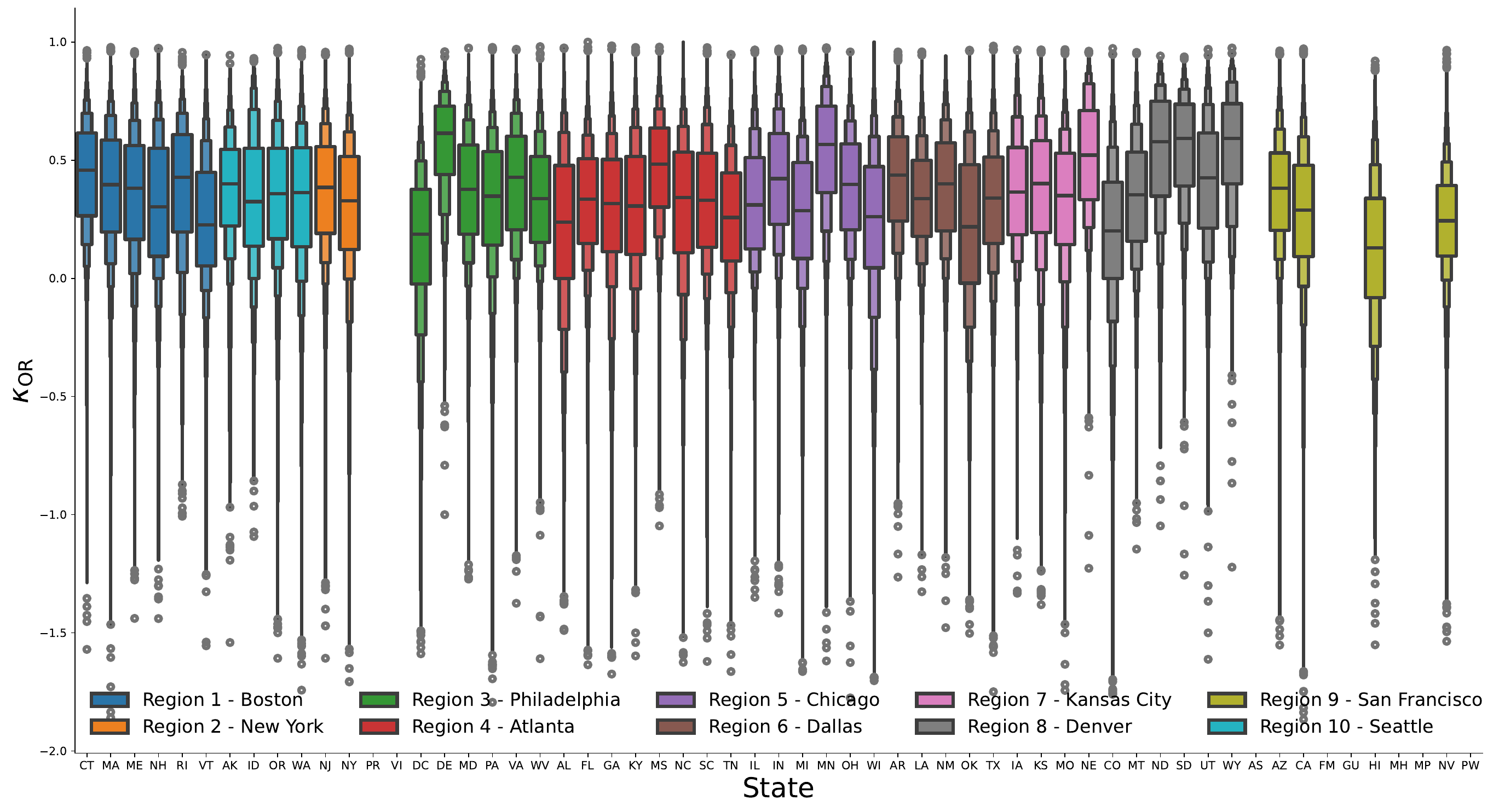}
    \caption{\textbf{Regional Distribution of Ollivier-Ricci Curvature} This figure displays the regional differences in care delivery structure as measured by $\ORC$. We plot curvature values for each edge (provider connection) for all 2017 \PRN, separated by state and colored by region. In comparison with $\FRC$ (see \Cref{fig:forman-distribution}), we see much smaller variance in $\ORC$ values which is expected based on their construction in \Cref{eq:forman} and \Cref{eq:orc}. Although $\ORC$ is known to be a more \emph{expressive} measure for capturing fine-grained network structure, although it can be much more expensive to compute for large networks.}
    \label{fig:orc-distribution}
\end{figure}

\begin{figure}[hb!]
    \centering
    \begin{subfigure}[b]{0.45\textwidth}
        \centering
        \includegraphics[width=\textwidth]{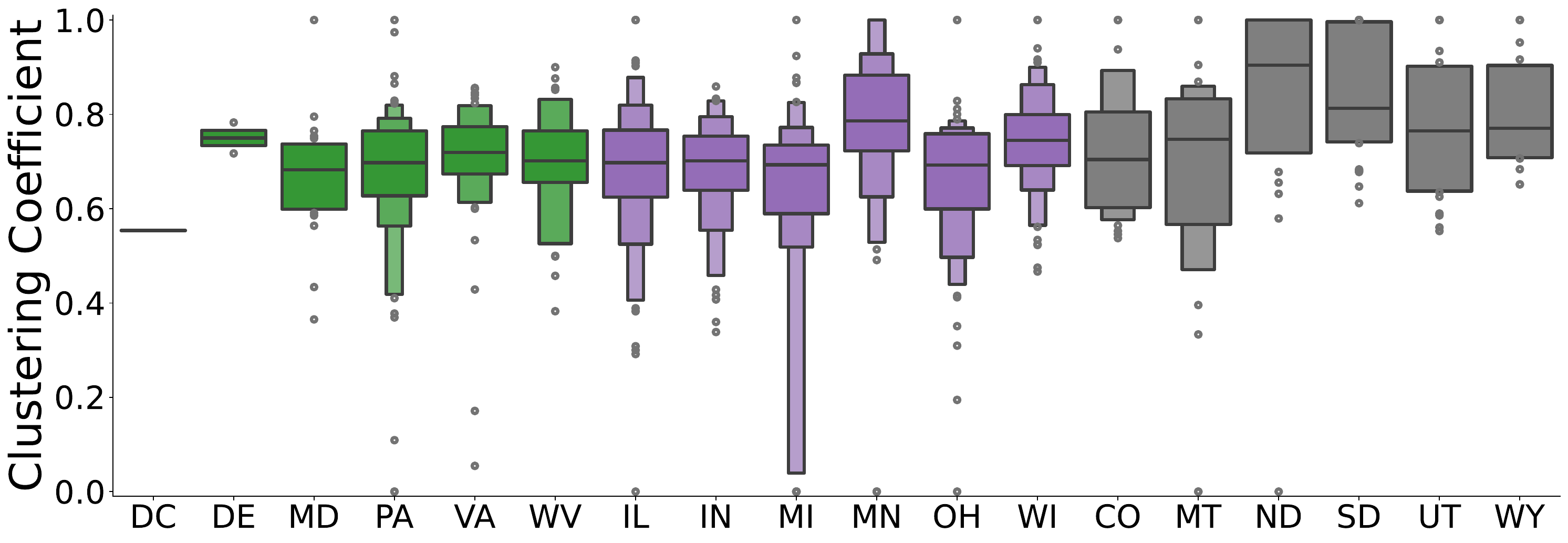}
        \caption{Clustering coefficients}
        \label{fig:cluster_coeff}
    \end{subfigure}
    \hfill
    \begin{subfigure}[b]{0.45\textwidth}
        \centering
        \includegraphics[width=\textwidth]{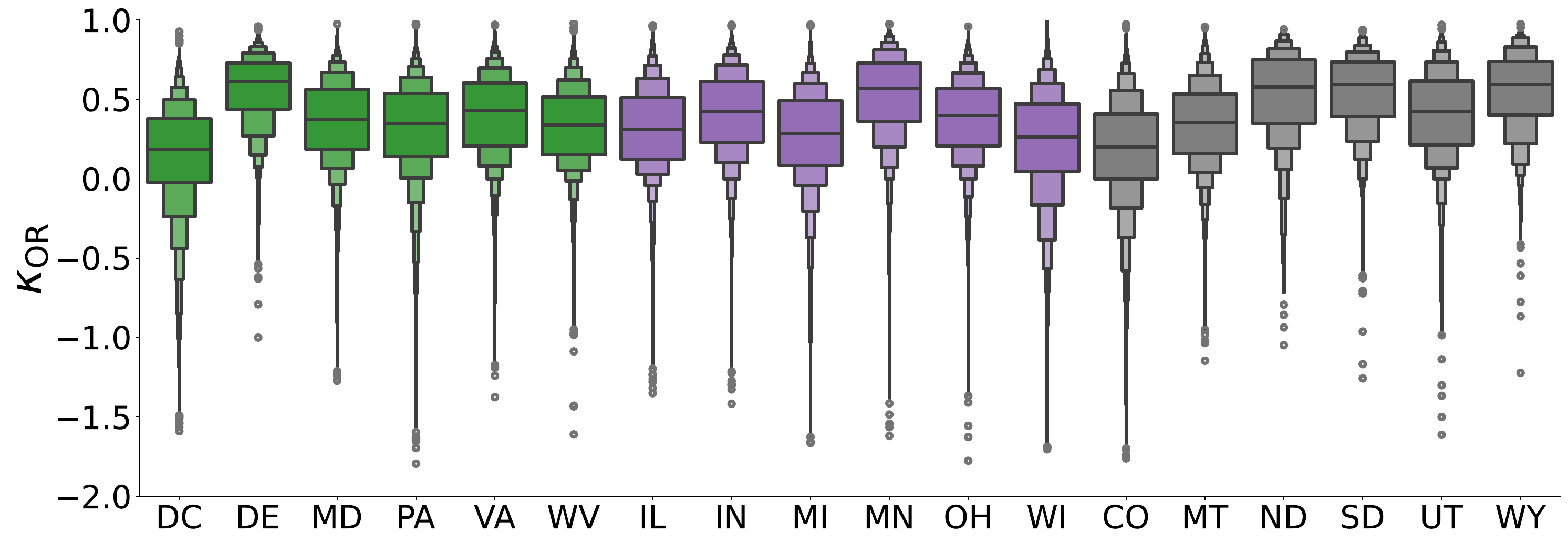}
        \caption{Ollivier-Ricci Edge Curvatures}
        \label{fig:orc}
    \end{subfigure}
    \caption{\textbf{Comparing Standard Network Features and Ricci Curvature}. For a select set of states in the Philadelphia (green), Chicago (purple), and Denver (gray) HSA regions, we visualize distributions of \emph{clustering coefficients} (CC) in comparison to $\ORC$. These are different measures (with clearly different distributions) for capturing connectivity in a network. In the context of care delivery positive CC and $\ORC$ values together could indicate high-functioning subregions of a Physician Referral Network.}
    \label{fig:selected-cluster-orc}
\end{figure}

Additionally, we provide distributions of Forman-Ricci and Ollivier-Ricci Curvature values for network edges, grouped by state and region in \Cref{fig:forman-distribution,fig:orc-distribution}. As we observe from the distribution, ORC values are constrained to a much tighter range, with the most negatively curved edges in our distribution approaching a value of $-2$, while all edge curvatures are bounded above by $+1$.
In comparison, Forman-Ricci, due to its dependence on node degree and the number of triangles, can produce very large curvature values, both negative and positive. Given the known trade-off between locality, expressivity, and efficiency between Forman-Ricci and Ollivier-Ricci Curvature \cite{southern_curvature_2023}, further analysis is required to understand which descriptors are most useful for specific healthcare management tasks. Despite clear agreement between various classical network science features (e.g, \Cref{fig:selected-cluster-orc}), Ricci curvature measures are known to be much more expressive in capturing network structure. Moreover, there is precedence for alleviating bottlenecks in graph neural networks by rewiring negatively curved edges in a network \cite{topping2022understanding}. This provides a novel way to potentially ``rewire''  care delivery networks using these measures in order to improve healthcare quality in regions that suffer from care fragmentation.

\begin{figure}[ht!]
    \centering
    \includegraphics[width=\textwidth]{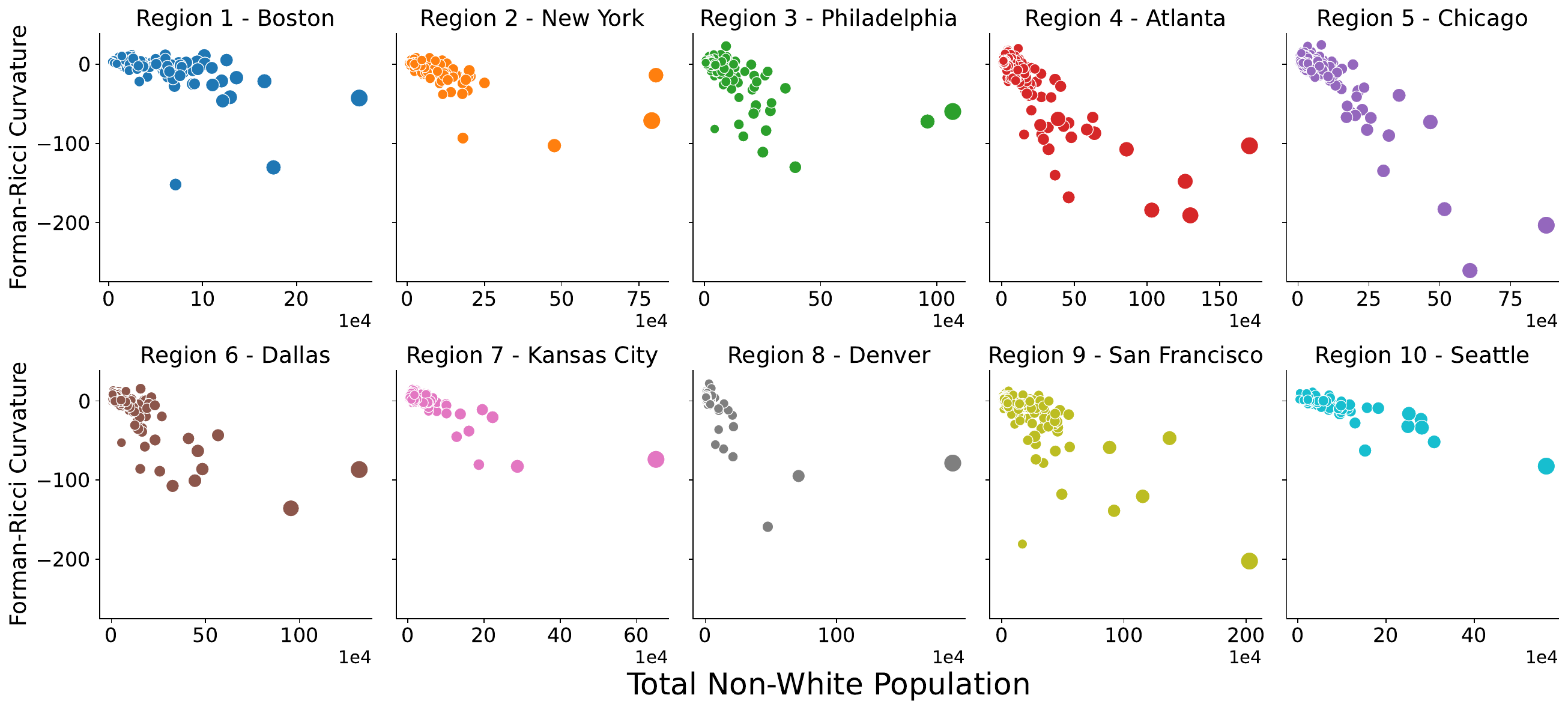}
    \caption{\textbf{Ricci-Curvature vs. Non-White Population}. Across different healthcare regions, we visualize the correlation between average Forman-Ricci ($\FRC$) curvature and total non-white population (node sizes correspond to network size). Despite clear inter-regional variation, there is consistent correlation between high non-white population in large networks with highly negative average curvature. Although these results are preliminary, they are consistent with prior work suggesting that \PRN serving historically marginalized populations exhibit systematically less robust connectivity and therefore are less equipped to deliver high quality, low cost care.}
    \label{fig:curvature_vs_demo}
\end{figure}

As part of our preliminary analysis, we compared Forman-Ricci Curvature with various population statistics, including Medicare enrollment, total mortality, and the proportion of non-white populations. As seen in \Cref{fig:curvature_vs_demo}, our findings reveal that large networks exhibiting low average curvature—indicating many potential bottlenecks in care delivery—correlate with high Medicare enrollment, elevated mortality rates, and higher proportions of Black and Hispanic populations. These results (which can be reproduced interactively with \apparent) suggest that negatively curved edges are indicative of structural properties in physician networks commonly found in less affluent HSAs.

Our analysis demonstrates significant variation in curvature across different regions, effectively capturing disparities in network structures. The observed negative curvature in areas with higher minority populations highlights the potential for developing recommendation technologies that could enhance patient flow and improve care accessibility with minimal intervention.
However, much remains to be explored, and we believe that the tools and data used in this study provide promising future avenues to better understand and address the complexities within \PRN.

\section{Discussion}
\subsection{Significance for Healthcare Management}
Our open-source infrastructure enables network analysis on medical claims data. This opens the door for exciting advancements in healthcare management. By leveraging network analysis techniques, we can uncover intricate patterns and relationships within healthcare delivery systems that were previously hidden. This is crucial for several reasons:

\begin{compactitem}
    \item \textbf{Optimization of Care Coordination.} Understanding referral patterns and physician collaborations can help identify inefficiencies and bottlenecks in patient flow, leading to more effective care coordination.
    \item \textbf{Resource Allocation.} Network analysis can inform strategic decisions regarding resource distribution, ensuring that healthcare facilities and services are optimally positioned to meet patient needs.
    \item \textbf{Policy Development.} Insights from network structures can guide policymakers in developing initiatives that foster more robust and integrated healthcare systems.
\end{compactitem}
By collecting and combining standard network features, local census data, healthcare efficacy, treatment outcomes, along with Ricci curvature, we hope to empower researchers in healthcare management to make expedited progress on unraveling and correcting care fragmentation across the United States. 
\subsection{The Promise of Ricci Curvature}

Ricci curvature measures for characterizing edges in a network have proven to be expressive both in terms of structural analysis of networks and identifying bottlenecks in information flow. Our work is the first to bring these geometric measures into the field of healthcare management. In particular, we compute and openly distribute our results of computing \emph{Ollivier-Ricci} and \emph{Forman-Ricci} curvature, and present promising results that indicate that bottlenecks in \PRN (as measured by curvature) correlate highly with metadata associated with healthcare systems that are known to receive a lower standard of care.

\subsection{Exploration Prototype}

To further the impact of our research, we provide an open-access dataset and analysis suite, \apparent~that combines medical claims data with various demographic and socioeconomic factors. This provides access to network features and abundant metadata, along with prototypes for exploring/visualizing user queries. Additionally, we provide a Python repository\footnote{Our codebase is maintained at: \url{https://github.com/aidos-lab/apparent}} that allows users to interact directly with the dataset. This repository supports building, customizing, and reproducing \PRN and associated structural features, as well as visualizing networks and feature distributions. By offering access to curvature features and other network metrics, we empower both technical and non-technical researchers to explore the data, fostering a collaborative environment for discovering new insights.

\subsection{Limitations}

While our approach offers many advantages, there are several limitations to consider:

\begin{compactitem}
    \item \textbf{Computational Tractability.} Calculating $\ORC$ on massive networks is computationally intensive, limiting its scalability. Thus, we recommend users to prioritize $\FRC$ for studying large scale \PRN.
    \item \textbf{Network Specifications.} We only build and analyze unweighted, undirected networks, which may overlook the complexities of real-world healthcare interactions. We hope to extend our analyses to more complex network constructions in the future.
    \item \textbf{Feature Completeness.} Our prototype may lack some critical features necessary for a comprehensive understanding of healthcare management. This prototype serves as a foundation for future enhancements, and can easily be updated as necessary.
\end{compactitem}

\subsection{Future Work}

Looking ahead, we aim to expand our research in several key areas:

\begin{compactitem}
    \item \textbf{Network Evolution.} Studying how healthcare networks evolve over time to identify trends and predict future changes.
    \item \textbf{Thorough Data Analysis.} Conducting a more detailed analysis of existing data to derive actionable insights that can influence policy and management decisions.
    \item \textbf{Regional Characteristics.} Understanding the distinctive features of regional healthcare networks and how they relate to specific policies and practices.
    \item \textbf{Edge Prediction}. Developing methods for predicting new referrals that could alleviate bottlenecks in patient flow, utilizing negative curvature as a predictive metric.
\end{compactitem}

We hope \apparent~will serve as a valuable resource, enabling researchers and practitioners to explore networks, network features (including curvature), and various other metadata, inspiring deeper investigations and support data-driven decision-making in healthcare management.

\section*{Acknowledgements}
We would like to acknowledge the authors and developers of the \href{https://datasette.io/}{\texttt{Datasette}} package, whose work has been instrumental in the development of our open-source prototype. Their tool has enabled us to efficiently manage, explore, and visualize our dataset, significantly enhancing our ability to analyze and share our findings. We also thank the members of the CARE-CONNECT team, as well as Corinna Coupette for her valuable insights during the early stages of the project and her assistance in creating visualizations, and Katharina Limbeck for her helpful review of the final manuscript. Research reported in this publication was supported by the National Heart, Lung, And Blood Institute of the National Institutes of Health under Award Number R01HL167816. The content is solely the responsibility of the authors and does not necessarily represent the official views of the National Institutes of Health.
B.R.\ is partially supported by the Bavarian state government with
funds from the \emph{Hightech Agenda Bavaria}. 

%
%

\bibliographystyle{splncs04}
\bibliography{references}

\begin{thebibliography}{10}
\providecommand{\url}[1]{\texttt{#1}}
\providecommand{\urlprefix}{URL }
\providecommand{\doi}[1]{https://doi.org/#1}

\bibitem{agha2019fragmented}
Agha, L., Frandsen, B., Rebitzer, J.B.: Fragmented division of labor and
  healthcare costs: Evidence from moves across regions. Journal of Public
  Economics  \textbf{169},  144--159 (2019)

\bibitem{an2018analysis}
An, C., O'Malley, A.J., Rockmore, D.N., Stock, C.D.: Analysis of the {US}
  patient referral network. Statistics in Medicine  \textbf{37}(5),  847--866
  (2018)

\bibitem{an2018referral}
An, C., O’Malley, A.J., Rockmore, D.N.: Referral paths in the {US} physician
  network. Applied Network Science  \textbf{3},  1--24 (2018)

\bibitem{barnett2012physician}
Barnett, M.L., Christakis, N.A., O’Malley, J., Onnela, J.P., Keating, N.L.,
  Landon, B.E.: Physician patient-sharing networks and the cost and intensity
  of care in {US} hospitals. Medical Care  \textbf{50}(2),  152--160 (2012)

\bibitem{barnett2011mapping}
Barnett, M.L., Landon, B.E., O'malley, A.J., Keating, N.L., Christakis, N.A.:
  Mapping physician networks with self-reported and administrative data. Health
  Services Research  \textbf{46}(5),  1592--1609 (2011)

\bibitem{casalino2015physician}
Casalino, L.P., Pesko, M.F., Ryan, A.M., Nyweide, D.J., Iwashyna, T.J., Sun,
  X., Mendelsohn, J., Moody, J.: Physician networks and ambulatory
  care-sensitive admissions. Medical care  \textbf{53}(6),  534--541 (2015)

\bibitem{cebul2008organizational}
Cebul, R.D., Rebitzer, J.B., Taylor, L.J., Votruba, M.E.: Organizational
  fragmentation and care quality in the {US} healthcare system. Journal of
  Economic Perspectives  \textbf{22}(4),  93--113 (2008)

\bibitem{CMS2439F}
{Centers for Medicare \& Medicaid Services}: Medicaid program; establishing
  minimum standards for medicaid managed care plans. Federal Register (May
  2024), docket Number: CMS-2439-F, RIN: 0938-AU99

\bibitem{coupette2023ollivierricci}
Coupette, C., Dalleiger, S., Rieck, B.: {O}llivier-{R}icci curvature for
  hypergraphs: A unified framework. In: International Conference on Learning
  Representations (2023), \url{https://openreview.net/forum?id=sPCKNl5qDps}

\bibitem{Devriendt_2022}
Devriendt, K., Lambiotte, R.: Discrete curvature on graphs from the effective
  resistance. Journal of Physics: Complexity  \textbf{3}(2),  025008 (2022).
  \doi{10.1088/2632-072x/ac730d}

\bibitem{dugoff_scoping_2018}
DuGoff, E.H., Fernandes-Taylor, S., Weissman, G.E., Huntley, J.H., Pollack,
  C.E.: A scoping review of patient-sharing network studies using
  administrative data. Translational Behavioral Medicine  \textbf{8}(4),
  598--625 (2018). \doi{10.1093/tbm/ibx015}

\bibitem{everson2022electronic}
Everson, J., Adler-Milstein, J.: Electronic connectivity among {US} hospitals
  treating shared patients. Medical Care  \textbf{60}(12),  880--887 (2022)

\bibitem{everson2018repeated}
Everson, J., Funk, R.J., Kaufman, S.R., Owen-Smith, J., Nallamothu, B.K.,
  Pagani, F.D., Hollingsworth, J.M.: Repeated, close physician coronary artery
  bypass grafting teams associated with greater teamwork. Health Services
  Research  \textbf{53}(2),  1025--1041 (2018)

\bibitem{frandsen2015care}
Frandsen, B.R., Joynt, K.E., Rebitzer, J.B., Jha, A.K.: Care fragmentation,
  quality, and costs among chronically ill patients. American Journal of
  Managed Care  \textbf{21}(5),  355--362 (2015)

\bibitem{funk2018association}
Funk, R.J., Owen-Smith, J., Kaufman, S.A., Nallamothu, B.K., Hollingsworth,
  J.M.: Association of informal clinical integration of physicians with cardiac
  surgery payments. JAMA Surgery  \textbf{153}(5),  446--453 (2018)

\bibitem{gandre_care_2020}
Gandré, C., Beauguitte, L., Lolivier, A., Coldefy, M.: Care coordination for
  severe mental health disorders: an analysis of healthcare provider
  patient-sharing networks and their association with quality of care in a
  {French} region. BMC Health Services Research  \textbf{20}(1), ~548 (2020).
  \doi{10.1186/s12913-020-05173-x}

\bibitem{gebhart2021go}
Gebhart, T., Fu, X., Funk, R.J.: Go with the flow? a large-scale analysis of
  health care delivery networks in the {U}nited {S}tates using {H}odge theory.
  In: IEEE International Conference on Big Data. pp. 3812--3823 (2021)

\bibitem{ghomrawi_physician_2018}
Ghomrawi, H.M.K., Funk, R.J., Parks, M.L., Owen-Smith, J., Hollingsworth, J.M.:
  Physician referral patterns and racial disparities in total hip replacement:
  {A} network analysis approach. PLOS One  \textbf{13}(2),  e0193014 (2018).
  \doi{10.1371/journal.pone.0193014}

\bibitem{graves2023physician}
Graves, J.A., Lee, D., Leszinsky, L., Nshuti, L., Nikpay, S., Richards, M.,
  Buntin, M.B., Polsky, D.: Physician patient sharing relationships within
  insurance plan networks. Health services research  \textbf{58}(5),
  1056--1065 (2023)

\bibitem{hollingsworth2015differences}
Hollingsworth, J.M., Funk, R.J., Garrison, S.A., Owen-Smith, J., Kaufman, S.R.,
  Landon, B.E., Birkmeyer, J.D.: Differences between physician social networks
  for cardiac surgery serving communities with high versus low proportions of
  black residents. Medical care  \textbf{53}(2),  160--167 (2015)

\bibitem{javalgi1993physicians}
Javalgi, R., Joseph, W.B., Gombeski~Jr, W.R., Lester, J.A.: How physicians make
  referrals. Journal of Health Care Marketing  \textbf{13}(2) (1993)

\bibitem{juo2019care}
Juo, Y.Y., Sanaiha, Y., Khrucharoen, U., Chang, B.H., Dutson, E., Benharash,
  P.: Care fragmentation is associated with increased short-term mortality
  during postoperative readmissions: a systematic review and meta-analysis.
  Surgery  \textbf{165}(3),  501--509 (2019)

\bibitem{kim2019informal}
Kim, D., Funk, R.J., Yan, P., Nallamothu, B.K., Zaheer, A., Hollingsworth,
  J.M.: Informal clinical integration in {M}edicare accountable care
  organizations and mortality following coronary artery bypass graft surgery.
  Medical Care  \textbf{57}(3),  194--201 (2019)

\bibitem{kim2023structure}
Kim, K.D., Funk, R.J., Zaheer, A.: Structure in context: A morphological view
  of whole network performance. Social Networks  \textbf{72},  165--182 (2023)

\bibitem{landon_variation_2012}
Landon, B.E., Keating, N.L., Barnett, M.L., Onnela, J.P., Paul, S., O’Malley,
  A.J., Keegan, T., Christakis, N.A.: Variation in {Patient}-{Sharing}
  {Networks} of {Physicians} {Across} the {United} {States}. JAMA
  \textbf{308}(3),  265--273 (2012). \doi{10.1001/jama.2012.7615}

\bibitem{landon2018patient}
Landon, B.E., Keating, N.L., Onnela, J.P., Zaslavsky, A.M., Christakis, N.A.,
  O’Malley, A.J.: Patient-sharing networks of physicians and health care
  utilization and spending among {M}edicare beneficiaries. JAMA Internal
  Medicine  \textbf{178}(1),  66--73 (2018)

\bibitem{matthews2023within}
Matthews, L.J., Damberg, C.L., Zhang, S., Escarce, J.J., Gibson, C.B., Schuler,
  M., Popescu, I.: Within-physician differences in patient sharing between
  primary care physicians and cardiologists who treat white and black patients
  with heart disease. Journal of the American Heart Association
  \textbf{12}(22),  e030653 (2023)

\bibitem{ni_ricci_2015}
Ni, C.C., Lin, Y.Y., Gao, J., Gu, X.D., Saucan, E.: Ricci {Curvature} of the
  {Internet} {Topology} (2015). \doi{10.48550/arXiv.1501.04138}

\bibitem{pollack2013patient}
Pollack, C.E., Weissman, G.E., Lemke, K.W., Hussey, P.S., Weiner, J.P.: Patient
  sharing among physicians and costs of care: a network analytic approach to
  care coordination using claims data. Journal of General Internal Medicine
  \textbf{28},  459--465 (2013)

\bibitem{popescu2024segregation}
Popescu, I., Gibson, B., Matthews, L., Zhang, S., Escarce, J.J., Schuler, M.,
  Damberg, C.L.: The segregation of physician networks providing care to black
  and white patients with heart disease: Concepts, measures, and empirical
  evaluation. Social Science \& Medicine  \textbf{343},  116511 (2024)

\bibitem{shortell1974determinants}
Shortell, S.M.: Determinants of physician referral rates: an exchange theory
  approach. Medical Care  \textbf{12}(1),  13--31 (1974)

\bibitem{shortell1971physician}
Shortell, S.M., Anderson, O.W.: The physician referral process: a theoretical
  perspective. Health Services Research  \textbf{6}(1), ~39 (1971)

\bibitem{sia_ollivier-ricci_2019}
Sia, J., Jonckheere, E., Bogdan, P.: Ollivier-{Ricci} {Curvature}-{Based}
  {Method} to {Community} {Detection} in {Complex} {Networks}. Scientific
  Reports  \textbf{9}, ~9800 (2019). \doi{10.1038/s41598-019-46079-x}

\bibitem{snow2020patient}
Snow, K., Galaviz, K., Turbow, S.: Patient outcomes following interhospital
  care fragmentation: a systematic review. Journal of General Internal Medicine
   \textbf{35},  1550--1558 (2020)

\bibitem{southern_curvature_2023}
Southern, J., Wayland, J., Bronstein, M.M., Rieck, B.: Curvature {Filtrations}
  for {Graph} {Generative} {Model} {Evaluation}. In: International Conference
  on Learning Representations (2023),
  \url{https://openreview.net/forum?id=Dt71xKyabn}

\bibitem{southern_expressive_2023}
Southern, J., Wayland, J., Bronstein, M.M., Rieck, B.: On the {Expressive}
  {Power} of {Ollivier}-{Ricci} {Curvature} on {Graphs}  (2023),
  \url{https://openreview.net/forum?id=F1fuuUYui1}

\bibitem{topping2022understanding}
Topping, J., Giovanni, F.D., Chamberlain, B.P., Dong, X., Bronstein, M.M.:
  Understanding over-squashing and bottlenecks on graphs via curvature. In:
  International Conference on Learning Representations (2022),
  \url{https://openreview.net/forum?id=7UmjRGzp-A}

\end{thebibliography}

\end{document}